\documentclass[preprint,12pt]{elsarticle}

\usepackage{amsmath,amssymb}
\usepackage{booktabs}
\usepackage{graphicx}
\usepackage{adjustbox}
\usepackage{hyperref}

\journal{The North American Journal of Economics and Finance}

\begin{document}

\begin{frontmatter}

\title{Synchronization Reveals Market Structure That Classical Randomness Tests Cannot Detect}

\author[uvm,bq]{Nam H. Le\corref{cor1}}
\ead{nam.le@uvm.edu}
\cortext[cor1]{Corresponding author.}
\address[uvm]{Department of Computer Science, University of Vermont, Burlington, VT 05405, USA}
\address[bq]{BunnyQuant Capital, Burlington, VT, USA}

\begin{abstract}
Whether asset prices contain structure beyond a random walk is
classically assessed with variance-ratio tests of the random-walk
hypothesis. Synchronization measures from physics, notably the Kuramoto
order parameter applied across trading timescales, have been proposed as
an alternative diagnostic of market state, but they have been validated
on single asset classes and never compared with the variance-ratio test
on the same data. We compute a cross-timeframe Kuramoto coherence measure
and a heteroskedasticity-robust Lo--MacKinlay variance ratio on identical
windows for 61 instruments spanning cryptocurrencies, foreign exchange,
metals, equity indices and single-name equities. The two diagnostics are
statistically independent. We explain this analytically: coherence
decomposes exactly into a direction-agreement channel and a
volatility-agreement channel, and under sign-symmetric returns the
direction channel is independent of the variance ratio and of every other
statistic built from squared returns; simulations confirm the prediction.
Coherence tracks the directional cleanliness of recent price moves in all
five asset classes, whereas the variance-ratio test cannot reject the
random walk for most of the panel. Held to finance benchmarks with strict
information timing, however, coherence adds almost nothing to
heterogeneous autoregressive volatility forecasts out of sample, and
trading rules built on it earn no gross return outside cryptocurrency,
where trading the coherence direction at high coherence breaks even at
about five basis points per round trip. Failing to reject weak-form
efficiency therefore does not rule out directional structure, but the
structure that synchronization reveals is mostly descriptive and
economically thin.
\end{abstract}


\end{frontmatter}

\section{Introduction}
\label{sec:intro}

Whether asset prices follow a random walk is among the oldest empirical
questions in financial economics. Weak-form market efficiency implies that
past prices carry no exploitable information about future prices
\citep{fama1970}, and the variance-ratio test of Lo and MacKinlay
\citep{lomackinlay1988}, together with its multiple-comparison and
bootstrap refinements \citep{chowdenning1993,kim2006}, has become the
standard short-sample tool for checking it \citep{charles2009}. The
evidence it has produced is mixed and varies across markets and over time
\citep{limbrooks2011}, consistent with the view that efficiency is an
evolving property of a market rather than a fixed one \citep{lo2004}. At
the same time, directional regularities such as time-series momentum are
documented across many asset classes \citep{moskowitz2012}, which raises a
practical question for anyone who relies on a randomness test: does a
failure to reject the random walk mean that there is no directional
structure to find, or only that the test is not looking for it?

A separate line of work, originating in physics, measures market state in
a different way. Synchronization, the spontaneous alignment of rhythm
among weakly coupled oscillators, is described by the Kuramoto order
parameter, which tracks how coherently a population of phase oscillators
moves together, rising toward one as the population locks into phase and
falling toward zero as it desynchronizes \citep{kuramoto1975,pikovsky2001}.
The same order parameter has proved a working diagnostic of collective
state in systems that are not literally oscillators, from power grids,
where loss of synchrony precedes cascading blackouts \citep{dorfler2013},
to neural, biological and engineered networks \citep{dorfler2014}.
Financial markets have been proposed as another such system: price series
observed across multiple timescales have been modeled as a population of
coupled oscillators, with the Kuramoto order parameter used to forecast
turning points in market indices \citep{fioriti2012} and, more recently,
to explain pink-noise-like fluctuation statistics in economic time series
through synchronization and amplitude demodulation \citep{morikawa2026}.
These applications report that the coherence measure carries real
information about market state.

Two gaps remain in this line of work, however. First, each of these
applications has been validated on a narrow slice of market data,
typically a single index or a single asset class, leaving open whether the
measured coherence is a property of markets in general or an artifact of
the one dataset examined. Second, and more fundamentally, none of this
work has been checked against the diagnostic that financial econometrics
already uses to answer the same underlying question: is there exploitable
structure in the price series, or is it indistinguishable from a random
walk? The variance-ratio test of Lo and MacKinlay \citep{lomackinlay1988}
is the discipline's own short-sample-valid answer to that question, and to
our knowledge it has never been run alongside a Kuramoto-style coherence
measure on the same data to see whether the two diagnostics agree.

This matters beyond a single methodological checkbox. If a physics-derived
synchronization measure and a field-native statistical test disagree, that
disagreement is informative: either one of the two is not measuring what
it claims to measure, or they are legitimately sensitive to different
layers of structure in the same data. We address both gaps directly. We
apply the same Kuramoto-style coherence measure and the same
variance-ratio test, identically constructed, to 61 instruments across
five asset classes -- cryptocurrency, foreign exchange, metals, equity
indices, and single-name equities -- and we report, for the first time,
their joint behavior on the same price windows. We find that the two
diagnostics are statistically independent of each other, that the
coherence measure's descriptive content generalizes across every asset
class we test while its forecasting content largely does not, and that the
variance-ratio test leaves the large majority of this panel
indistinguishable from a random walk. We then go one step beyond
measurement: we derive an exact decomposition of the coherence measure
into a direction channel and a volatility channel
(Section~\ref{sec:methods-decomposition}), prove that the direction
channel is independent of the variance ratio under a mild symmetry
assumption on returns, and confirm by controlled simulation
(Section~\ref{sec:results-mechanism}) that the observed independence is a
structural property of the measurement pair itself -- one that persists
even on synthetic processes with heavy, dialed-in persistence -- rather
than a contingent property of the markets studied. Finally, we hold the
coherence measure to the benchmarks a finance reader would apply. With
information timing handled strictly, coherence adds almost nothing to
heterogeneous autoregressive volatility forecasts out of sample, and
simple trading rules built on it earn no gross return outside
cryptocurrency; in cryptocurrency, trading in the direction of the
coherence phase when coherence is high earns a small gross return whose
break-even cost lies between maker and taker fees. Together, these
results indicate that a synchronization-based diagnostic imported from
physics is not simply restating what a classical statistical test of
market efficiency already says, but that the structure it sees is mostly
descriptive: a failure to reject the random walk leaves room for
directional structure, yet that structure is economically thin. Both
halves of this conclusion matter before either tool is used,
individually or jointly, to make claims about markets.

\section{Related work}
\label{sec:related}

The Kuramoto order parameter's reach beyond its original
chemical-oscillator setting is well documented: beyond the power-grid
application already noted \citep{dorfler2013}, phase-coupled oscillator
models describe collective rhythm formation across neural, biological,
and engineered networks generally \citep{dorfler2014}, and the broader
synchronization framework was assembled into a single unified treatment
spanning physics, chemistry, and biology by Pikovsky, Rosenblum and Kurths
\citep{pikovsky2001}. What these applications share is not a literal
population of oscillators but a consistent bookkeeping of phase agreement
that tracks a real collective state, a property that transfers whenever a
system can be given a well-defined phase at each observed scale.

Importing diagnostics from physics into finance has, more broadly, its
own three-decade history under the name econophysics, formalized by
Mantegna and Stanley's treatment of correlations and complexity in price
series \citep{mantegna1999}. A canonical result from this tradition is
the application of random matrix theory to the empirical correlation
matrix of stock returns: the bulk of the eigenvalue spectrum of a
correlation matrix built from hundreds of assets matches the spectrum
predicted for a matrix of pure noise, isolating only a handful of
eigenvalues as carrying genuine market-wide structure
\citep{laloux1999}. This and related work established a template this
paper follows: take a tool built for a different kind of system, measure
it carefully on market data, and report exactly how much of the signal
survives a null-model comparison. What that template has not yet been
applied to is a synchronization-based diagnostic benchmarked against the
discipline's own classical efficiency test, rather than against an
internally-generated null model.

Within this tradition, two applications of the Kuramoto order parameter
itself to market data illustrate both its promise and its narrow
validation to date. Fioriti and Chinnici \citep{fioriti2012} treat a
single market index as a population of oscillators with natural
frequencies estimated from Hilbert and Fourier analysis of the index
itself, and report that the resulting order parameter anticipates turning
points in that index. Morikawa, Morikawa and Nakamichi
\citep{morikawa2026} take the model in a different direction, showing
that a stochastic extension of the Kuramoto dynamics reproduces the
pink-noise-like (1/f) fluctuation spectrum widely observed in economic
time series, via a mechanism of synchronization and amplitude
demodulation rather than ad hoc noise assumptions. Both results are
genuine contributions, but both are also demonstrations on a single
series or a single class of series, with no attempt to establish whether
the measured coherence is a property of markets broadly or a property of
the one index examined.

On the other side of the comparison this paper draws, the random-walk
hypothesis has its own long lineage, independent of and predating any
physics import. Fama's review \citep{fama1970} organized market
efficiency into the weak, semi-strong, and strong forms still used today,
with the weak form, the claim that price histories contain no exploitable
structure, as the one a diagnostic like ours is most directly positioned
against. The variance-ratio test of Lo and MacKinlay
\citep{lomackinlay1988} gave this claim a short-sample-valid statistical
test, and Chow and Denning \citep{chowdenning1993} extended it to a joint
test across multiple aggregation horizons simultaneously, controlling the
overall false-positive rate rather than testing each horizon in
isolation, the same multiple-horizon discipline we apply in
Section~\ref{sec:methods-vr}. Rank- and sign-based versions
\citep{wright2000} and wild-bootstrap versions \citep{kim2006} improve the
test's behaviour under heteroskedastic and non-normal returns, and the
resulting literature is reviewed by Charles and Darn\'e
\citep{charles2009}. Applied across markets and decades, these tests give
a mixed and time-varying verdict \citep{limbrooks2011}, which the adaptive
markets hypothesis \citep{lo2004} interprets as efficiency evolving with
market conditions rather than holding or failing once and for all.
Cryptocurrency markets have become a frequent test case: early Bitcoin
prices failed a battery of efficiency tests \citep{urquhart2016}, and
cryptocurrency returns display strong time-series momentum and are
predicted by measures of investor attention \citep{liu2021}.

Any claim that a new diagnostic forecasts the size of future price moves
has to be measured against the strong existing benchmarks for
volatility forecasting. Conditional-variance models \citep{bollerslev1986}
and realized-volatility models built from intraday data
\citep{andersenbollerslev1998,abdl2003} forecast volatility well, and the
heterogeneous autoregressive model of realized volatility
\citep{corsi2009}, motivated by traders operating at different horizons
\citep{muller1997}, captures its long memory with three simple averages.
Because realized volatility is itself an imperfect proxy, forecast
comparisons should use loss functions that are robust to proxy noise
\citep{patton2011}, and out-of-sample comparisons of nested models need
tests that account for the extra parameter \citep{clarkwest2007}. In
return forecasting, the out-of-sample record of most proposed predictors
is weak relative to the historical mean
\citep{welchgoyal2008,campbell2008}, and the number of factors tried in
the literature calls for higher significance hurdles
\citep{harvey2016}. A multi-timeframe coherence measure is, in this
light, one more candidate predictor that must show value beyond these
benchmarks, which Section~\ref{sec:results-forecast} tests directly.

The coherence measure is also built from technical regime labels, which
connects it to the literature on technical analysis and regimes. Simple
moving-average and trading-range rules carried information about
subsequent returns in long samples of US equities
\citep{brock1992}, and pattern-based technical signals can be given a
statistical foundation \citep{lomamaysky2000}. Trend-following
strategies earn positive average returns across asset classes over long
histories \citep{moskowitz2012,hurst2017}, and scaling exposure by
forecast volatility improves risk-adjusted returns
\citep{moreira2017}. Regime-switching models \citep{hamilton1989} and
their applications in finance \citep{angtimmermann2012} provide the
statistical counterpart to the discrete regime labels that our measure
aggregates across timeframes.

Direct comparisons between a complexity-style diagnostic and a classical
efficiency test are not entirely unprecedented: Turvey \citep{turvey2007}
estimates the Hurst exponent via a scaled variance-ratio procedure and
finds it can diverge markedly from the same exponent estimated by
rescaled-range analysis on the same commodity price series, and
Shternshis, Mazzarisi and Marmi \citep{shternshis2022} measure Shannon
entropy directly against a market-efficiency null on high-frequency data,
finding a weak but genuine departure from efficiency. What neither of
these, nor any work we are aware of, has done is run a
phase-synchronization measure and a variance-ratio test side by side on
the same price windows to ask whether the two actually agree, the
comparison this paper reports for the first time, across five asset
classes rather than one.

Three gaps emerge from this literature, taken together. First, every
direct application of the Kuramoto order parameter to markets to date has
been validated on a single series or a single narrow class of series,
leaving its generality across asset classes untested. Second, the two
diagnostics this paper compares, phase synchronization and statistical
variance-ratio testing of the random-walk hypothesis, descend from
entirely separate literatures that have never been applied jointly to the
same data. Third, where complexity-style diagnostics have been checked
against classical efficiency tests at all, the comparison has been
limited to fractal (Hurst) and information-theoretic (entropy) measures,
never to a synchronization measure of this kind. Closing all three gaps
at once, on the same instruments and the same time windows, and then
holding the coherence measure to the forecasting and economic benchmarks
of the finance literature, is the contribution of the rest of this paper.

\section{Methods}
\label{sec:methods}

\subsection{Cross-timeframe phase coherence}
\label{sec:methods-coherence}

Consider an instrument observed on a ladder of $K$ nested timeframes
(e.g.\ 1m/5m/15m/1h, or 15m/1h/4h/1d; see Section~\ref{sec:data}). At each
timeframe $k \in \{1,\dots,K\}$, a window of recent bars is mapped to one
of four discrete regime labels. A concrete, fully public illustration of
how such a label can be constructed from price data alone is a technical
trend-volatility classifier: let $m_t$ be a 20-bar simple moving average
of the close price evaluated at the end of the window, with slope
$s = m_t - m_{t-5}$, and let $\sigma_t$ be the 10-bar rolling realized
volatility of log returns, with window median $\tilde\sigma$. The label is
\begin{equation}
  L_{\text{tech}} =
  \begin{cases}
    0 & s > 0,\ \sigma_t \leq \tilde\sigma \quad \text{(quiet uptrend)} \\
    1 & s > 0,\ \sigma_t > \tilde\sigma \quad \text{(volatile uptrend)} \\
    2 & s \leq 0,\ \sigma_t > \tilde\sigma \quad \text{(volatile downtrend)} \\
    3 & s \leq 0,\ \sigma_t \leq \tilde\sigma \quad \text{(quiet downtrend)}
  \end{cases}
  \label{eq:technical-classifier}
\end{equation}
so that quiet and volatile phases of the same trend direction are adjacent
in the label's natural cyclic order. This classifier is used exclusively
for a robustness check reported in Section~\ref{sec:discussion}, to test
whether the paper's central descriptive finding depends on the specific
classifier used. The classifier that generates the headline results in
Section~\ref{sec:results} follows the same general structure -- four
discrete regime states inferred from recent price behavior at each
timeframe -- but is a different, proprietary rule belonging to a deployed
commercial trading-analytics platform \citep{bunnyquant}, not disclosed
(Section~\ref{sec:discussion}).

Each label, from whichever classifier, is in turn mapped to a canonical
phase on the unit circle,
\begin{equation}
  \theta_k \in \Theta = \left\{0,\ \tfrac{\pi}{2},\ \pi,\ \tfrac{3\pi}{2}\right\},
  \label{eq:phase-alphabet}
\end{equation}
the four values equally spaced to represent a cyclical four-state process
with no preferred starting point. The assignment of labels to angles is
order-preserving with respect to the classifier's own regime sequence:
labels that the classifier transitions into one another, in its typical
operating sequence, are placed at adjacent quadrants, so that circular
distance between two $\theta$ values reflects transition proximity in the
underlying regime graph rather than an arbitrary indexing. This is the
only property of the label-to-angle map used anywhere in this paper.

Let $U \subseteq \{1,\dots,K\}$ denote the subset of timeframes for which
the classifier returns a usable label (excluded timeframes are counted but
never assigned an arbitrary phase, so that a missing reading is never
treated as evidence of neutrality), and let $w_k>0$ be a fixed
per-timeframe weight, longer timeframes weighted more heavily. Following
the mean-field reduction of the Kuramoto model \citep{kuramoto1975}, define
the complex order parameter
\begin{equation}
  z \equiv r\, e^{i\psi} = \frac{1}{W}\sum_{k \in U} w_k\, e^{i\theta_k},
  \qquad W = \sum_{k \in U} w_k, \qquad |U| \geq 2,
  \label{eq:kuramoto}
\end{equation}
with $r \in [0,1]$ the coherence magnitude and $\psi \in [0,2\pi)$ the mean
phase. By construction $r \to 1$ as the usable timeframes converge on a
single $\theta_k$ (fully synchronized) and $r \to 0$ as they spread
uniformly around the circle (fully desynchronized); the circular variance
$1-r$ is the standard measure of dispersion for angular data.

\begin{figure}[htbp]
\centering
\includegraphics[width=0.95\linewidth]{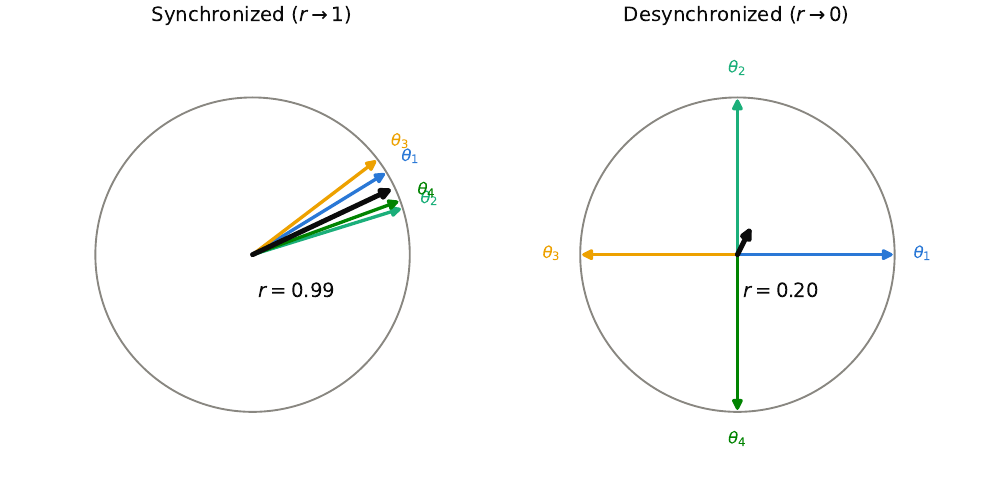}
\caption{The order parameter of Equation~\ref{eq:kuramoto} as a vector
sum. Left: four timeframe phasors clustered together add constructively,
giving a long resultant ($r$ near 1). Right: phasors at the four canonical
angles $\Theta$ of Equation~\ref{eq:phase-alphabet} cancel, giving a short
resultant ($r$ near 0). Thin arrows: individual $\theta_k$; thick black
arrow: the resultant $z$.}
\label{fig:kuramoto-schematic}
\end{figure}

A naive
directional statistic follows directly from the imaginary part of $z$,
\begin{equation}
  \beta \equiv r \sin\psi = \mathrm{Im}(z),
  \label{eq:bias}
\end{equation}
projecting the mean phase onto a fixed bullish/bearish axis, scaled by how
strongly the timeframes agree; $\beta \in [-1,1]$.

As a baseline for comparison we also report a legacy linear-encoding
indicator from the same production system. Let $e: \Theta \to
\{-1,-\tfrac12,\tfrac12,1\}$ be an order-preserving encoding of the four
phases, and let $v_k = e(\theta_k)$ repeated $\lfloor 2w_k \rfloor$ times
for $k \in U$, concatenated into a single vector $v$. The legacy indicator
is
\begin{equation}
  L = \mathrm{clip}\!\left(1 - \frac{\mathrm{std}(v)}{1.5},\ 0,\ 1\right),
  \label{eq:legacy}
\end{equation}
a simpler dispersion-based measure of the same underlying
agreement-across-timeframes idea, predating the order-parameter
formulation in Equation~\ref{eq:kuramoto}.

\subsection{Variance-ratio test}
\label{sec:methods-vr}

For a log-return series $r_t$, $t=1,\dots,n$, with sample mean $\mu$, write
$x_t = r_t - \mu$. The variance ratio at aggregation horizon $q$ is
\begin{equation}
  VR(q) = \frac{\widehat{\sigma}^2_q}{q\,\widehat{\sigma}^2_1},
  \qquad
  \widehat{\sigma}^2_q = \frac{1}{m}\sum_{t=q}^{n}\Big(\sum_{j=0}^{q-1} x_{t-j}\Big)^{2},
  \qquad
  \widehat{\sigma}^2_1 = \frac{1}{n-1}\sum_{t=1}^n x_t^2,
  \label{eq:vr}
\end{equation}
with $m$ the appropriate unbiasing denominator for the overlapping
$q$-period sum. $VR(q)=1$ under the random-walk null, $VR(q)>1$ under
positive serial correlation (trending), and $VR(q)<1$ under mean
reversion. Because financial returns are heteroskedastic, we use the
robust statistic of Lo and MacKinlay \citep{lomackinlay1988} rather than
its homoskedastic counterpart, which rejects the random walk far too often
on real data for reasons unrelated to genuine structure. Writing
\begin{equation}
  \widehat\delta_j = \frac{n\sum_{t=j+1}^{n} x_t^2\, x_{t-j}^2}
                          {\left(\sum_{t=1}^n x_t^2\right)^2},
  \qquad
  \widehat\theta(q) = \sum_{j=1}^{q-1}\left[\frac{2(q-j)}{q}\right]^2 \widehat\delta_j,
  \label{eq:lm-variance}
\end{equation}
the heteroskedasticity-robust test statistic is
\begin{equation}
  z(q) = \frac{\sqrt{n}\left(VR(q)-1\right)}{\sqrt{\widehat\theta(q)}}
  \ \xrightarrow{d}\ \mathcal{N}(0,1) \quad \text{under } H_0.
  \label{eq:lm-z}
\end{equation}
We evaluate $z(q)$ at $q \in \{2,4,8,16\}$ and, following the joint-test
logic of Chow and Denning \citep{chowdenning1993}, reject the random-walk
null only if $\min_q p(q) < \alpha / M$ with $M=4$ the number of horizons
tested, a Bonferroni correction that controls the overall false-positive
rate against the alternative of testing each horizon in isolation and
reporting the most extreme one. Fitting the log-linear regression
\begin{equation}
  \log VR(q) = (2H-1)\log q + \varepsilon_q
  \label{eq:hurst-fit}
\end{equation}
by ordinary least squares across the same four horizons gives an implied
Hurst exponent $H = 0.5 + \widehat{\text{slope}}/2$, consistent with the
self-similarity scaling $VR(q)=q^{2H-1}$ of a fractional process, so the
estimator stays in Hurst units without losing the fractal framing.

\begin{figure}[htbp]
\centering
\includegraphics[width=0.78\linewidth]{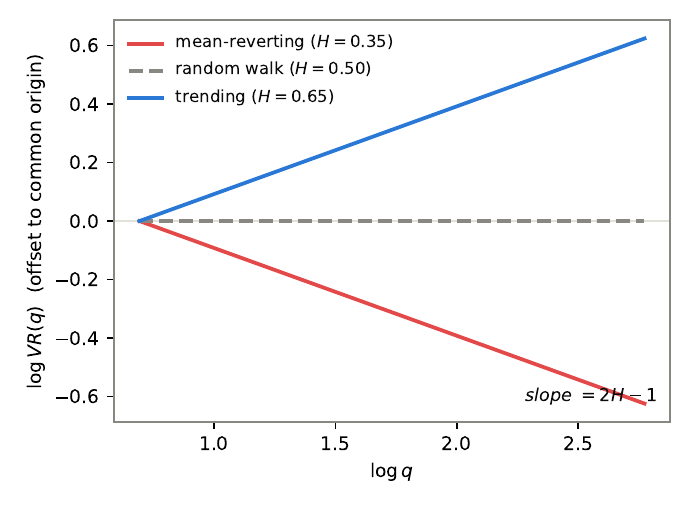}
\caption{Geometry of Equation~\ref{eq:hurst-fit}: $\log VR(q)$ against
$\log q$ for three idealized processes. The fitted slope is $2H-1$, so a
random walk ($H=0.5$) is flat, a trending process ($H>0.5$) slopes up, and
a mean-reverting process ($H<0.5$) slopes down.}
\label{fig:vr-hurst-schematic}
\end{figure}

This estimator was calibrated before being applied to any market data (see
Section~\ref{sec:validation}): on 400 trials of pure Gaussian white noise at
each of five sample sizes ($n \in \{120, 180, 250, 500, 1000\}$), it falsely
rejects $H_0$ between 4.5\% and 6.2\% of the time against a nominal 5\%
level, and recovers a planted fractional Gaussian noise Hurst exponent with
bias under 0.06 across the tested range $H \in \{0.35, 0.45, 0.50, 0.60,
0.70\}$.

\subsection{An exact decomposition and a symmetry theorem}
\label{sec:methods-decomposition}

The two constructions above are not merely different formulas; they live
in different function spaces of the return path, and making that precise
yields a theorem that predicts -- rather than merely describes -- the
central empirical result of Section~\ref{sec:results-joint}.

Write $X=(x_1,\dots,x_n)$ for the log-return vector of one instrument.
Every quantity in Section~\ref{sec:methods-vr} -- $VR(q)$, the robust
statistic $z(q)$, the implied Hurst exponent $\hat H$, realized volatility
-- is built from \emph{products} of returns and is therefore an
\emph{even} functional of the path: $Q(-X)=Q(X)$. The coherence measure is
not. To see exactly how, note that any label in the four-state alphabet of
Equation~\ref{eq:phase-alphabet} factors into two binary coordinates: a
direction bit $u=\mathbb 1[L\in\{0,1\}]$ and a volatility bit
$h=\mathbb 1[L\in\{1,2\}]$ (for the public classifier of
Equation~\ref{eq:technical-classifier} these are literally
$u=\mathbb 1[s>0]$ and $h=\mathbb 1[\sigma_t>\tilde\sigma]$). Writing
$d=2u-1$ and $v=2h-1$ for the corresponding signs, direct evaluation of
the four cases gives $\cos\theta=(d-v)/2$ and $\sin\theta=(d+v)/2$.
Substituting into Equation~\ref{eq:kuramoto} with
$\bar D = \frac{1}{W}\sum_{k\in U} w_k d_k$ and
$\bar V = \frac{1}{W}\sum_{k\in U} w_k v_k$ yields, exactly,
\begin{equation}
  r^2 = \frac{\bar D^2 + \bar V^2}{2},
  \qquad
  \beta = \frac{\bar D + \bar V}{2}:
  \label{eq:decomposition}
\end{equation}
the squared order parameter splits, in quadrature, into a
cross-timeframe \emph{direction-agreement} channel $\bar D$ and a
cross-timeframe \emph{volatility-agreement} channel $\bar V$. This
identity holds for any classifier mapped to the phase alphabet; what is
classifier-specific is the character of the two channels as functionals
of $X$. For the public classifier of
Equation~\ref{eq:technical-classifier}, each direction bit is the sign of
a fixed \emph{linear} filter of the path ($s=c^\top X$ for a
moving-average-difference weight vector $c$), making $\bar D$ an
\emph{odd}-symmetric statistic, while each volatility bit is a functional
of squared returns, making $\bar V$ even -- the same parity class as the
variance ratio itself.

This parity split has an exact consequence.

\textbf{Proposition (direction--variance-ratio independence).} Assume the
return vector is sign-symmetric, $X \stackrel{d}{=} -X$ -- true for any
zero-mean Gaussian process, including conditionally heteroskedastic ones
(e.g.\ GARCH with symmetric innovations), and violated only by genuine
skewness or leverage asymmetry. Let $s=c^\top X$ be any linear statistic
with $c\neq 0$ and let $Q(X)$ be any even functional. Then
$\mathrm{sign}(s)$ is \emph{independent} of $Q(X)$:
$P\!\left(\mathrm{sign}(s)=1 \mid Q(X)\right)=\tfrac12$ almost surely,
hence $\mathrm{Cov}\!\left(\mathrm{sign}(s),f(Q(X))\right)=0$ for every
function $f$.

\emph{Proof.} Under $X \mapsto -X$, $\mathrm{sign}(c^\top X)$ flips sign
while $Q(X)$ is invariant, so
$(\mathrm{sign}(s),\,Q(X)) \stackrel{d}{=} (-\mathrm{sign}(s),\,Q(X))$.
Since $\mathrm{sign}(s)\in\{-1,+1\}$ almost surely, the conditional law
of $\mathrm{sign}(s)$ given $Q(X)$ is forced to be the fair coin,
which is the claimed independence. \hfill$\square$

Applied to this paper's pair of diagnostics: under sign-symmetric
returns, each direction bit $d_k$ -- and any statistic of the collection
$\{d_k\}$ alone, including $\bar D$ -- is exactly independent of $VR(q)$,
of $\hat H$, and of realized volatility, with no approximation and no
appeal to sample size. Two things follow. First, the naive directional
statistic $\beta$ of Equation~\ref{eq:bias} contains $\bar D$ as its odd
half, so its empirical emptiness across every panel
(Section~\ref{sec:results-descriptive}) is not a fact about markets that
happened to come out that way; it is forced by the algebra whenever
returns are approximately sign-symmetric. Second, the theorem deliberately
does \emph{not} cover the volatility channel: $\bar V$ is even, the same
parity as the variance ratio, so nothing in the symmetry argument
prevents $\bar V$ (and through it $r$) from correlating with $\hat H$.
Whatever relationship survives between $r$ and the variance ratio must
travel through that one channel -- a sharp, measurable prediction that
Section~\ref{sec:results-mechanism} tests on controlled synthetic
processes where the ground truth is known.

\subsection{Data and asset-class panel}
\label{sec:methods-data}
\label{sec:data}

All data were pulled to local disk once (no live API dependency during
analysis) across three sources, summarized in Table~\ref{tab:data}. Each
source has its own native format and sampling quirks (exchange-specific
kline encodings, broker tick-volume conventions, provider-specific
calendar gaps); a per-source retrieval module normalizes all of them into
one common schema, OHLCV bars on a UTC-indexed, deduplicated, sorted
timestamp axis, before any measurement in Section~\ref{sec:methods} is
applied, so that the coherence and variance-ratio pipelines are entirely
source-agnostic downstream of ingestion.

\begin{table}[htbp]
\centering
\caption{Asset-class panel: source, instrument count, timeframe ladder.}
\label{tab:data}
\begin{adjustbox}{max width=\textwidth}
\begin{tabular}{@{}lllll@{}}
\toprule
Asset class & Source & $n$ & Ladder & Base \\
\midrule
Cryptocurrency & Binance & 24 & 15m/1h/4h/1d & 1h \\
FX majors+crosses & Broker CFD (MT5) & 10 & 1m/5m/15m/1h & 5m \\
Metals & Broker CFD (MT5) & 2 & 1m/5m/15m/1h & 5m \\
Equity indices & Broker CFD (MT5) & 5 & 1m/5m/15m/1h & 5m \\
Single-name equities & Broker CFD (MT5) & 6 & 1m/5m/15m/1h & 5m \\
Cryptocurrency (cross-check) & Broker CFD (MT5) & 5 & 1m/5m/15m/1h & 5m \\
Futures proxies & yfinance & 4 & 1h/4h/1d & 4h \\
Single-name equities (daily) & yfinance & 6 & 1h/4h/1d & 4h \\
\bottomrule
\end{tabular}
\end{adjustbox}
\end{table}

Cryptocurrency appears twice in Table~\ref{tab:data}, sourced
independently through Binance and through a retail broker's CFD feed, so
that the descriptive finding of Section~\ref{sec:results-descriptive} can
be checked against venue-specific quoting and liquidity artifacts rather
than taken on the strength of a single exchange; the two panels agree in
sign and rough magnitude (Section~\ref{sec:results-descriptive}). Two
cryptocurrency CFD symbols originally planned for the broker panel
(ADAUSD, DOGEUSD) were dropped: the broker discontinued quoting them after
August 2024, so no data exists in the requested window -- not a
methodological failure, and not a loss of coverage, since both assets are
already present in the Binance panel.

The rolling lookback window used by the regime classifier (1000 bars) is a
window measured in \emph{bars}, not calendar time, and the
ladder-construction code treats it identically regardless of what
timeframe those bars represent. This means a ``fast'' ladder
(1m/5m/15m/1h, needing only weeks of history to accumulate 1000+ one-minute
bars) and a ``slow'' ladder (15m/1h/4h/1d, needing years) are testing the
\emph{same statistical construction} at genuinely different absolute
timescales -- not a shortcut taken because deep intraday history was
unavailable. Where a source's available history was short relative to this
window (notably yfinance's intraday data, capped at 725 days by the
provider), the per-symbol lookback and sampling step were adjusted downward
proportionally (never below a 300-bar floor) rather than silently producing
an under-powered result; this is logged per instrument and does not change
the measured relationships, only the density of sampling.

Six of the eight panels above are, to our knowledge, the first time this
specific coherence construction has been measured outside a single
cryptocurrency exchange.

\subsection{Validation design}
\label{sec:validation}

\textbf{Permutation control.} For the cryptocurrency panel, each series was
randomly shuffled (preserving the marginal return distribution, destroying
temporal structure) and the same measurement pipeline re-run. A real effect
should disappear or reverse under shuffling; an artifact of the bookkeeping
should survive or strengthen.


\textbf{Volatility confound.} Because both diagnostics could in principle be
re-expressions of simple realized volatility, we additionally check the
rank correlation between the coherence measure and trailing realized
volatility within each panel (reported in Table~\ref{tab:descriptive}).
Values stay small (median Spearman correlation between $-0.13$ and $+0.18$
across the eight panels), indicating the coherence measure is not simply
volatility measured a second way.

\textbf{Variance-ratio estimator calibration.} As described in
Section~\ref{sec:methods-vr}, the variance-ratio estimator was validated
against simulated white noise and simulated fractal Gaussian noise with
known Hurst exponents \emph{before} being applied to any of the market data
in this paper, establishing its false-positive rate and recovery bias
independently of any market-data result.

\textbf{Mechanism simulation.} To test the predictions of
Section~\ref{sec:methods-decomposition} where the ground truth is known,
the full measurement pipeline -- the public classifier of
Equation~\ref{eq:technical-classifier}, the order parameter in its
decomposed form (Equation~\ref{eq:decomposition}, verified algebraically
identical to Equation~\ref{eq:kuramoto}), and the identical variance-ratio
estimator -- is run end-to-end on synthetic price paths from four
controlled generating processes: (i) i.i.d.\ Gaussian returns (pure null);
(ii) GARCH(1,1) at two strengths (null for the variance ratio -- zero
return autocorrelation -- but with genuine volatility clustering, to
isolate whether clustering alone induces any coupling); (iii) AR(1) with
$\varphi \in [-0.15, 0.20]$ (dialed-in linear persistence); and (iv)
fractional Gaussian noise with
$H_{\text{true}} \in \{0.35,\dots,0.65\}$, generated by the same
Davies--Harte routine used in the calibration above. A synthetic
four-rung ladder (aggregation factors 1/4/16/96, mirroring the
15m/1h/4h/1d cryptocurrency ladder, with the same per-rung weights) is
sampled per path, with a 120-bar per-rung classifier lookback and the
same 250-bar causal variance-ratio window as the empirical analysis.
Eighteen cells (3 null + 8 AR(1) + 7 fGn) of 80 paths each are measured;
within each cell, correlations between the coherence quantities
($r$, $\bar D$, $\bar V$) and $\hat H$ are computed \emph{per path} and
summarized by the median and an across-path sign test -- the same
per-instrument aggregation rule used for the real panel in
Section~\ref{sec:results-joint}, avoiding the pseudo-replication that
pooling autocorrelated windows from one path would introduce.

\subsection{Information timing, forecast baselines and economic value}
\label{sec:methods-econ}

\textbf{Information timing.} All price bars in the panel are time-stamped
at their opening time. A higher-timeframe label enters the coherence
measure at a base-timeframe bar only if the higher-timeframe bar has
\emph{closed} by the close of that base bar; for example, on the
15m/1h/4h/1d ladder the daily label used at an hourly bar is the label of
the previous completed day. Aligning bars on their opening times instead, an
easy mistake when bars are stamped at the open, exposes the coherence
measure to up to one higher-timeframe bar of future prices. Every result
reported in this paper uses the corrected alignment, and
Section~\ref{sec:results-forecast} reports how much opening-time
alignment would inflate the forecast statistics.

\textbf{Out-of-sample magnitude forecasts.} To ask whether coherence adds
anything to what standard volatility forecasts already know, the target is
the log high--low range over the next $H$ base-timeframe bars, a range-based
proxy for realized volatility \citep{parkinson1980}, with the log of
future realized volatility as a robustness target. Two nested baselines are
estimated. The first is a heterogeneous autoregressive (HAR) specification
in the spirit of Corsi \citep{corsi2009} and the heterogeneous-market
view of M\"uller et al.\ \citep{muller1997}: log realized volatility over the trailing 6, 24 and 120
base bars. The second adds the log trailing 24-bar range, the log 14-bar
average true range and the 24-bar efficiency ratio, so that coherence has
to beat everything a practitioner would already compute from the same
bars. Forecast origins are non-overlapping (every $H$ bars). Each model is
estimated by OLS on an expanding window, using the first half of each
instrument's sample for initial estimation and re-estimating at ten
equally spaced points over the evaluation half. We report the
out-of-sample $R^2$ of the model with coherence added relative to the
baseline \citep{campbell2008} and the Clark--West test \citep{clarkwest2007} for
nested models (one-sided, 5\%), per instrument, aggregated by panel.

\textbf{Economic value.} Economic significance is measured with simple
rules whose positions use only information available at the forecast
origin. The benchmark rules are a long-only position (to separate any
signal from drift) and time-series momentum, the sign of the trailing
24-bar return \citep{moskowitz2012}. The coherence rules take the sign of
the directional bias $\beta$ (Equation~\ref{eq:bias}), unconditionally and
split by whether $r$ is above or below the 2/3 quantile of its own
\emph{expanding} past distribution. An efficiency-ratio filter with the
same quantile rule serves as a non-coherence baseline. Each rule holds its
position for $H$ bars from non-overlapping origins placed on a clock grid
common to the instruments of a panel. Returns are averaged with equal
weight across the instruments active at each origin, so that one time
series per panel and rule is tested; this absorbs cross-sectional
correlation (cryptocurrencies, in particular, co-move strongly) instead of
counting correlated instruments as independent evidence. Rather than fix a
single transaction cost, we report gross returns per trade, the
break-even round-trip cost, and annualized Sharpe ratios on a cost
frontier of 0.5, 2, 5 and 15 basis points per round trip, which brackets
the range from exchange-traded index futures (well under one basis point
of notional per contract round trip) through cryptocurrency perpetual
futures at maker rates and retail contracts for difference to
cryptocurrency spot trading at taker rates. Because the proprietary
classifier cannot be released, forecast and economic evaluations use the
public classifier of Equation~\ref{eq:technical-classifier}, so that every
number in Sections~\ref{sec:results-forecast} and~\ref{sec:results-econ}
can be reproduced from public data and the equations in this paper.

\section{Results}
\label{sec:results}

We state the central result first, then show by controlled simulation
that it is the structural property predicted by
Section~\ref{sec:methods-decomposition}, and finally build up the two
pieces of evidence that make it meaningful rather than vacuous.

\subsection{The central result: coherence and randomness are independent}
\label{sec:results-joint}

\begin{figure}[htbp]
\centering
\includegraphics[width=0.72\linewidth]{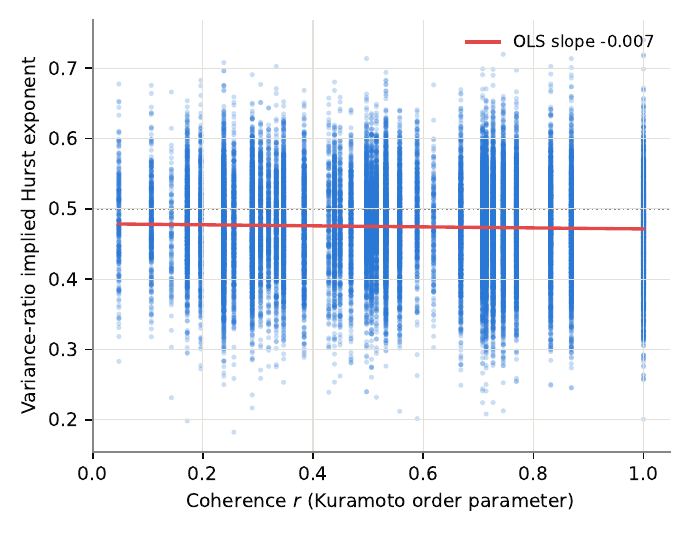}
\caption{Coherence $r$ against variance-ratio-implied Hurst exponent,
cryptocurrency (Binance) panel, $n=24$ instruments pooled. The near-flat
OLS fit is the visual counterpart of Table~\ref{tab:joint}'s small slope:
a statistically detectable but practically negligible relationship.}
\label{fig:joint-scatter}
\end{figure}

If phase coherence and the variance ratio were just two ways of asking
``is this market trending,'' they would move together. They do not. We
correlate the coherence measure $r$ against the variance-ratio-implied
Hurst exponent, computed causally on the same trailing 250-bar window,
timestamp-aligned within each panel (Table~\ref{tab:joint},
Figure~\ref{fig:joint-scatter}).

\begin{table}[htbp]
\centering
\caption{Joint relationship between $r$ and VR-implied Hurst.}
\label{tab:joint}
\begin{adjustbox}{max width=\textwidth}
\begin{tabular}{@{}lrrrr@{}}
\toprule
Panel & $n$ & median $\rho(r,H)$ & sign-test $p$ & VR-reject rate, hi$-$lo $r$ \\
\midrule
Cryptocurrency (Binance) & 24 & $-0.028$ & $0.064$ (17/24 neg.) & $+0.000$ \\
FX                       & 10 & $-0.083$ & $0.021$ (9/10 neg.)  & $-0.004$ \\
Metals                   & 2  & $+0.090$ & $0.500$              & $+0.009$ \\
Equity indices            & 5  & $+0.012$ & $1.000$              & $-0.021$ \\
Single-name equities (fast) & 6 & $-0.035$ & $0.219$            & $-0.002$ \\
Cryptocurrency (MT5)      & 5  & $+0.004$ & $1.000$              & $+0.013$ \\
Futures proxies           & 4  & $-0.026$ & $1.000$              & $+0.006$ \\
Single-name equities (slow) & 6 & $-0.019$ & $0.688$            & $+0.003$ \\
\bottomrule
\end{tabular}
\end{adjustbox}
\end{table}

Only in FX ($n=10$) is the relationship significant by a sign test, with
Binance cryptocurrency ($n=24$) borderline ($p=0.064$), and the magnitude
is small enough everywhere (median correlations between $-0.08$ and
$+0.09$) to be of no practical use as a combined signal; the
high-$r$-minus-low-$r$ difference in VR rejection rate is close to zero in
every panel. We read this as a genuine, unanticipated finding rather than
a confound: phase coherence is sensitive to directional cleanliness as
seen through a discrete, path-dependent regime classifier, while the
variance ratio is sensitive to linear serial correlation in raw returns. A
market can score high on one and not the other, and in this panel it
typically does. \emph{Statistically detectable} here does not mean
\emph{practically meaningful} -- worth stating plainly, given how easy it
is to over-read a $p<0.01$ result.

This independence is informative only if each diagnostic is picking up
something real on its own. Section~\ref{sec:results-mechanism} first
establishes \emph{why} the independence holds; the remainder of this
section then establishes that each diagnostic is individually doing real
work.

\subsection{The independence is a property of the measurement pair, not
of the markets}
\label{sec:results-mechanism}

\begin{figure}[htbp]
\centering
\includegraphics[width=0.95\linewidth]{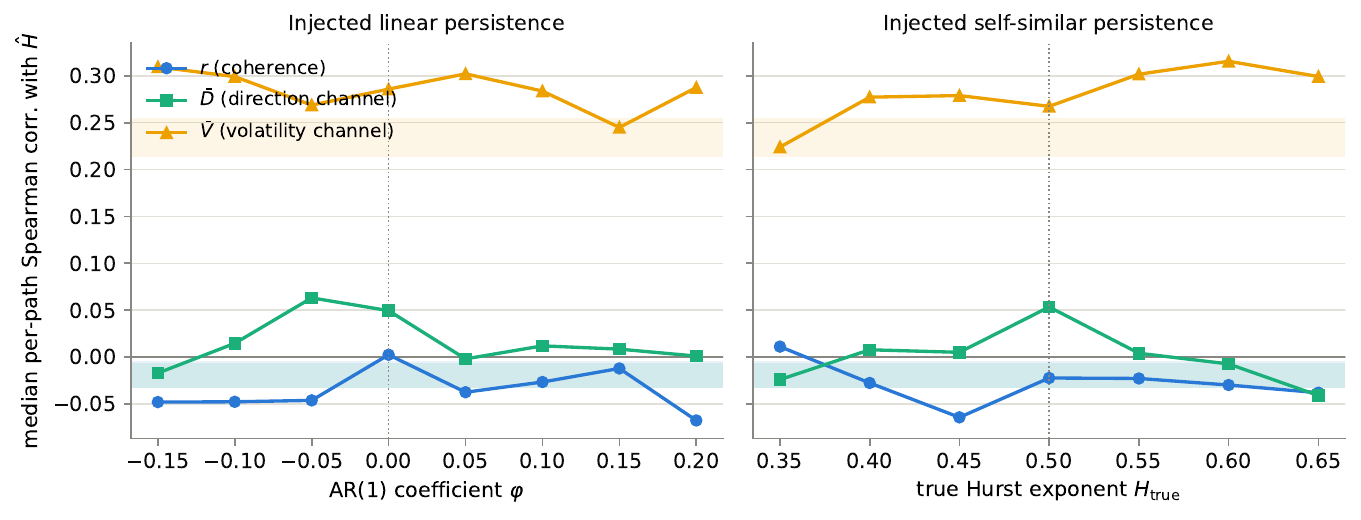}
\caption{Dose-response of the coherence--variance-ratio coupling on
synthetic paths with known ground truth. Median per-path Spearman
correlation of $r$, $\bar D$, and $\bar V$
(Equation~\ref{eq:decomposition}) with the VR-implied Hurst exponent
$\hat H$, against injected linear persistence (left: AR(1) coefficient
$\varphi$) and injected self-similar persistence (right: true Hurst
exponent of fractional Gaussian noise). Shaded bands: the range each
quantity occupies across the three persistence-free null cells
(i.i.d.\ and two GARCH strengths). No curve responds to the dose.}
\label{fig:mechanism-dose}
\end{figure}

The decomposition of Section~\ref{sec:methods-decomposition} predicts
that the direction channel $\bar D$ must be exactly uncoupled from
$\hat H$ under sign-symmetric returns, and leaves the volatility channel
$\bar V$ as the only route through which $r$ and the variance ratio could
legitimately co-move. The mechanism simulation of
Section~\ref{sec:validation} tests both statements -- and one further,
stronger question the theorem does not answer by itself: does injecting
\emph{genuine} persistence, which the variance ratio is built to detect,
finally couple the two diagnostics? Figure~\ref{fig:mechanism-dose}
summarizes all eighteen cells.

Three facts emerge. First, the theorem holds numerically everywhere it
should: the median per-path correlation between $\bar D$ and $\hat H$
stays within $[-0.041, +0.063]$ across all eighteen cells with no
consistent sign (smallest sign-test $p=0.057$) -- \emph{including} the
cells with heavy injected persistence, where the variance-ratio side has
abundant power (at $\varphi=0.20$ it rejects the random walk in 79\% of
windows, at $H_{\text{true}}=0.65$ in 91\%, and its mean $\hat H$ tracks
the planted truth from 0.35 to 0.62). This is the theorem working
exactly as proved: AR(1) and fractional Gaussian noise with Gaussian
innovations are still sign-symmetric processes, so persistence does not
open the direction channel; only distributional asymmetry can.

Second, the volatility channel carries a correlation that is real but
carries no structural information: $\bar V$ correlates with $\hat H$ at
$+0.21$ to $+0.32$ in every cell -- including pure i.i.d.\ noise, where
there is nothing to detect -- and the value does not move with
$\varphi$ or $H_{\text{true}}$. A coupling that is equally present under
white noise and completely insensitive to genuine persistence is a
finite-sample co-sensitivity of the two estimators to where the recent
volatility fluctuations happened to fall in the window, not a shared
reading of market structure.

Third, and most consequentially: the full coherence measure $r$ remains
uncoupled from $\hat H$ \emph{even when heavy genuine persistence is
injected} -- median per-path correlation between $-0.068$ and $+0.011$
across the entire sweep, with a weak tendency toward small negative
values at the dose extremes ($-0.068$ at $\varphi=0.20$, sign-test
$p=0.002$). That last pattern -- small, negative, significant only with
many units -- is quantitatively the same signature as the real panel in
Table~\ref{tab:joint} (medians between $-0.08$ and $+0.09$, significant
only in FX and borderline in the largest panel). The observed near-independence in real markets
therefore requires no market-specific explanation: it is what this
measurement pair produces on \emph{any} approximately sign-symmetric
return process, structured or not. We had initially conjectured the
opposite for the persistence cells -- that genuine self-similar structure
would couple the diagnostics through simultaneous cross-scale agreement
and lag correlation -- and the simulation falsified that conjecture;
we report this as stated rather than retrofitting the prediction,
because the falsification is precisely what upgrades the central result
of Section~\ref{sec:results-joint} from an empirical observation about
61 instruments to a structural property of the diagnostic pair.

\subsection{Coherence describes markets universally}
\label{sec:results-descriptive}

Table~\ref{tab:descriptive} reports the Spearman IC between $r$ and how
``clean'' the just-completed price move was, by panel, against a legacy
linear-encoding baseline and the naive directional-bias statistic.

\begin{table}[htbp]
\centering
\caption{Descriptive power (IC with past-move efficiency), by panel.
Asterisk: unanimous sign across every instrument in the panel.}
\label{tab:descriptive}
\begin{adjustbox}{max width=\textwidth}
\begin{tabular}{@{}lrrrrr@{}}
\toprule
Panel & $n$ & $r$ (coherence) & legacy & dir.\ bias & $r$ vs.\ vol \\
\midrule
Cryptocurrency (Binance)        & 24 & $+0.181^{*}$ & $+0.080$ & $-0.011$ & $+0.062$ \\
FX                               & 10 & $+0.199^{*}$ & $+0.086$ & $+0.005$ & $+0.086$ \\
Metals                          & 2  & $+0.261^{*}$ & $+0.170$ & $+0.069$ & $+0.107$ \\
Equity indices                   & 5  & $+0.202^{*}$ & $+0.085$ & $+0.059$ & $+0.102$ \\
Single-name equities (fast)      & 6  & $+0.216^{*}$ & $+0.119$ & $+0.011$ & $+0.159$ \\
Cryptocurrency (MT5)             & 5  & $+0.165^{*}$ & $+0.023$ & $+0.024$ & $+0.181$ \\
Futures proxies                  & 4  & $+0.249^{*}$ & $+0.145$ & $+0.200$ & $-0.133$ \\
Single-name equities (slow)      & 6  & $+0.376^{*}$ & $+0.179$ & $+0.049$ & $+0.029$ \\
\bottomrule
\end{tabular}
\end{adjustbox}
\end{table}

\begin{figure}[htbp]
\centering
\includegraphics[width=0.85\linewidth]{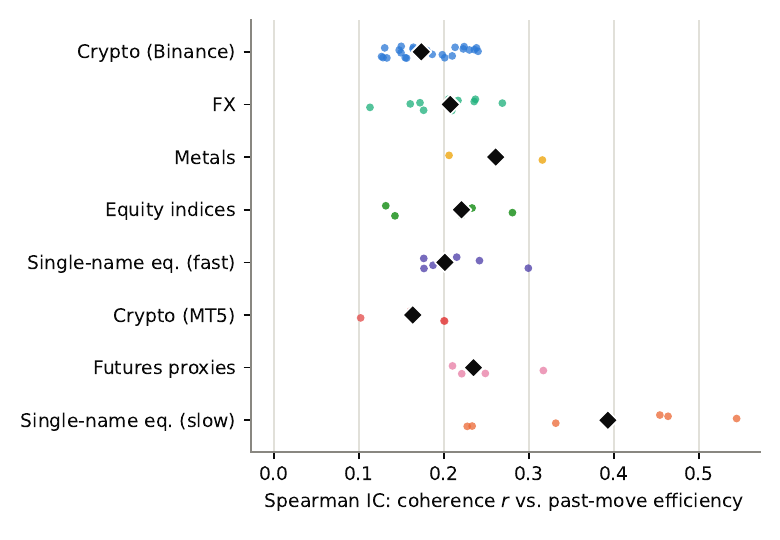}
\caption{Per-instrument descriptive IC (Spearman correlation between $r$
and past-move efficiency) by panel. Small points: individual instruments;
diamonds: panel median.}
\label{fig:descriptive-forest}
\end{figure}

Every panel is positive with unanimous sign, and every panel exceeds the
legacy baseline, by a factor of $1.5$--$2.4\times$ in seven panels and
about $7\times$ in the broker cryptocurrency panel. Figure~\ref{fig:descriptive-forest}
shows this is not an average hiding a few outliers: nearly every
individual instrument in every panel sits on the positive side of zero.

\subsection{Coherence adds little to standard volatility forecasts}
\label{sec:results-forecast}

Describing the recent past is not the same as forecasting the near
future, and a finance reader will rightly ask whether a coherence measure
tells a forecaster anything that realized volatility does not. Coherence
is positively rank-correlated with the subsequent high--low range in most
instruments (median IC $+0.03$ to $+0.05$ across horizons,
Table~\ref{tab:lookahead}), but range is strongly persistent, and the
relevant question is whether coherence adds to a volatility forecast that
already uses that persistence. Table~\ref{tab:forecast-oos} answers it
out of sample.

\begin{table}[htbp]
\centering
\caption{Out-of-sample value of coherence for forecasting the log future range, public classifier, corrected information timing. $\Delta R^2_{\text{OOS}}$: median across instruments of the out-of-sample $R^2$ of the model with $r$ added, relative to the baseline (in \%). CW: instruments with a significant Clark--West statistic (one-sided 5\%) out of instruments evaluated. HAR: trailing realized volatility over 6, 24 and 120 bars. HAR+: HAR plus trailing range, average true range and efficiency ratio.}
\label{tab:forecast-oos}
\small
\begin{adjustbox}{max width=\textwidth}
\begin{tabular}{@{}lrrrrrrrr@{}}
\toprule
 & \multicolumn{4}{c}{$H=6$} & \multicolumn{4}{c}{$H=24$} \\
\cmidrule(lr){2-5}\cmidrule(l){6-9}
Panel & $\Delta R^2$ HAR & CW & $\Delta R^2$ HAR+ & CW & $\Delta R^2$ HAR & CW & $\Delta R^2$ HAR+ & CW \\
\midrule
Cryptocurrency (Binance) & +0.41 & 22/24 & +0.27 & 20/24 & +0.42 & 12/24 & +0.17 & 6/24 \\
FX & +0.07 & 3/10 & -0.03 & 1/10 & -0.10 & 1/10 & -0.13 & 0/10 \\
Metals & -0.08 & 0/2 & -0.02 & 0/2 & -0.35 & 0/2 & -0.32 & 0/2 \\
Equity indices & -0.18 & 0/5 & -0.07 & 0/5 & -0.37 & 0/5 & -0.42 & 0/5 \\
Single-name equities (fast) & -0.30 & 0/6 & -0.16 & 0/6 & -0.65 & 1/6 & -0.68 & 1/6 \\
Cryptocurrency (MT5) & +0.21 & 3/5 & +0.02 & 1/5 & -0.23 & 1/5 & -0.18 & 1/5 \\
Futures proxies & -2.26 & 0/4 & -1.68 & 0/4 & -- & -- & -- & -- \\
Single-name equities (slow) & -- & -- & -- & -- & -- & -- & -- & -- \\
\midrule
All instruments & +0.18 & 28/56 & +0.01 & 22/56 & +0.09 & 15/52 & -0.02 & 8/52 \\
\bottomrule
\end{tabular}
\end{adjustbox}
\end{table}

Against the HAR baseline, adding coherence improves the out-of-sample
$R^2$ by a median of $0.18$ percentage points across all instruments at
the 6-bar horizon, significantly so by the Clark--West test for half of
them, and by $0.09$ points at 24 bars. Against the richer baseline, which
also contains trailing range, average true range and the efficiency ratio,
the median gain is zero at both horizons, and at 72 and 120 bars coherence
lowers out-of-sample accuracy against either baseline. The one panel
where coherence retains a consistent increment is Binance cryptocurrency:
$+0.41$ points over HAR and $+0.27$ points over the richer baseline at
6 bars, significant for 22 and 20 of the 24 coins. Using log future
realized volatility instead of range as the target gives the same
picture with even smaller gains (median $+0.05$ points over HAR and
$-0.03$ points over the richer baseline at 6 bars).
Repeating the comparison with the deployed classifier, against the
volatility controls stored with its output (trailing realized
volatility, average true range and past-window volatility), gives the
same pattern: median gains of $+0.25$ and $+0.38$ points in Binance
cryptocurrency at 6 and 24 bars (Clark--West significant for 12 of 24
coins at each), and negative median gains in every other panel. The
economically honest summary is that coherence is not a new volatility
predictor: whatever it knows about the size of the next move is almost
entirely already contained in trailing realized volatility and range,
with a small, short-horizon residual in continuously traded
cryptocurrency.

Table~\ref{tab:lookahead} shows why this check had to be run with the
corrected information timing of Section~\ref{sec:methods-econ}. Under
opening-time alignment, the median IC with future range is
$0.008$ to $0.024$ higher at every horizon (at $H=24$, $0.052$ against
$0.033$), because the higher-timeframe labels then contain part of the
very move being forecast. The bias is modest for the public classifier,
but it is always in the direction of overstating forecast power, and
multi-timeframe studies that do not state their alignment rule should be
read with this in mind.

\begin{table}[htbp]
\centering
\caption{Effect of information timing on forecast statistics, all instruments, public classifier. Median per-instrument Spearman IC between $r$ and the log future range, and median out-of-sample $R^2$ gain over HAR (\%), under the corrected alignment and under the earlier opening-time alignment.}
\label{tab:lookahead}
\begin{adjustbox}{max width=\textwidth}
\begin{tabular}{@{}lrrrr@{}}
\toprule
 & \multicolumn{2}{c}{IC with future range} & \multicolumn{2}{c}{$\Delta R^2_{\text{OOS}}$ over HAR} \\
\cmidrule(lr){2-3}\cmidrule(l){4-5}
Horizon & corrected & opening-time & corrected & opening-time \\
\midrule
$H=6$ & +0.046 & +0.054 & +0.18 & +0.20 \\
$H=24$ & +0.033 & +0.052 & +0.09 & +0.06 \\
$H=72$ & +0.046 & +0.057 & -0.16 & -0.27 \\
$H=120$ & +0.015 & +0.039 & -0.83 & -0.75 \\
\bottomrule
\end{tabular}
\end{adjustbox}
\end{table}

\subsection{Economic value}
\label{sec:results-econ}

Statistical forecast gains of a fraction of a percentage point of $R^2$
are unlikely to be worth trading, but the more natural economic use of a
coherence measure is directional: the order parameter's phase says which
way the timeframes agree, and $r$ says how strongly. Table~\ref{tab:econ}
reports equal-weight panel portfolios for the rules of
Section~\ref{sec:methods-econ}.

\begin{table}[htbp]
\centering
\caption{Economic value of coherence-based rules, equal-weight panel portfolios, public classifier, holding period $H$ base bars. Gross: mean return per holding period in basis points (equal to the break-even round-trip cost); $t$: $t$-statistic of the panel portfolio series; SR: annualized Sharpe ratio net of the stated round-trip cost (bps). Cryptocurrency (Binance), the only panel with a non-negligible coherence effect, and futures proxies are shown; all panels are in the replication files.}
\label{tab:econ}
\small
\begin{adjustbox}{max width=\textwidth}
\begin{tabular}{@{}llrrrrrr@{}}
\toprule
Panel, $H$ & Rule & Gross & $t$ & SR@0.5 & SR@2 & SR@5 & SR@15 \\
\midrule
Crypto (Binance), $H=6$ & Long only & -0.1 & -0.05 & -0.14 & -0.47 & -1.14 & -3.37 \\
 & Momentum & -1.3 & -0.69 & -0.50 & -0.93 & -1.78 & -4.63 \\
 & Coherence direction, all & +2.9 & +1.78 & +0.77 & +0.29 & -0.65 & -3.81 \\
 & Coherence direction, high $r$ & +4.7 & +2.14 & +1.00 & +0.64 & -0.06 & -2.41 \\
 & Coherence direction, low $r$ & +1.3 & +0.76 & +0.24 & -0.22 & -1.14 & -4.20 \\
\addlinespace
Crypto (Binance), $H=24$ & Long only & -0.3 & -0.03 & -0.05 & -0.13 & -0.30 & -0.86 \\
 & Momentum & +6.5 & +0.86 & +0.42 & +0.31 & +0.10 & -0.60 \\
 & Coherence direction, all & +9.2 & +1.36 & +0.68 & +0.56 & +0.33 & -0.46 \\
 & Coherence direction, high $r$ & +11.4 & +1.30 & +0.66 & +0.56 & +0.38 & -0.22 \\
 & Coherence direction, low $r$ & +9.0 & +1.33 & +0.66 & +0.55 & +0.31 & -0.47 \\
\addlinespace
Crypto (Binance), $H=72$ & Long only & +5.4 & +0.18 & +0.09 & +0.06 & +0.01 & -0.18 \\
 & Momentum & -4.4 & -0.19 & -0.12 & -0.15 & -0.22 & -0.46 \\
 & Coherence direction, all & +33.0 & +1.55 & +0.84 & +0.80 & +0.72 & +0.47 \\
 & Coherence direction, high $r$ & +24.4 & +0.92 & +0.49 & +0.46 & +0.40 & +0.19 \\
 & Coherence direction, low $r$ & +26.0 & +1.27 & +0.69 & +0.65 & +0.57 & +0.30 \\
\addlinespace
Futures proxies, $H=6$ & Long only & +6.7 & +1.22 & +1.06 & +0.80 & +0.29 & -1.41 \\
 & Momentum & +0.7 & +0.15 & +0.04 & -0.26 & -0.86 & -2.85 \\
 & Coherence direction, all & +2.4 & +0.56 & +0.41 & +0.08 & -0.58 & -2.80 \\
 & Coherence direction, high $r$ & +0.3 & +0.03 & -0.02 & -0.17 & -0.48 & -1.48 \\
 & Coherence direction, low $r$ & -0.1 & -0.03 & -0.13 & -0.42 & -1.01 & -2.96 \\
\bottomrule
\end{tabular}
\end{adjustbox}
\end{table}

Outside cryptocurrency, no coherence-based rule earns a gross return
distinguishable from zero in any panel at any horizon, so the question of
transaction costs does not arise: even exchange-traded index futures,
whose round-trip cost is well below one basis point of notional, cannot
make a zero gross edge profitable. In the futures-proxy panel the
largest gross return belongs to the long-only benchmark, which reflects
equity drift over the sample rather than any timing skill.

Binance cryptocurrency is the exception. Trading in the direction of the
coherence phase earns $+2.9$ basis points per 6-hour holding period
unconditionally and $+4.7$ basis points ($t=2.14$ on the panel portfolio)
when $r$ is in the top third of its own past distribution, against
$+1.3$ basis points when it is not; long-only and simple momentum earn
nothing at this horizon. The break-even round-trip cost of the high-$r$
rule is therefore about 5 basis points, and its annualized Sharpe ratio is
$1.00$ at 0.5 basis points and $0.64$ at 2 basis points, costs within
reach of maker execution on cryptocurrency perpetual futures, but
negative at the 15 basis points of retail spot trading at taker rates. At
24 and 72 hours, the coherence direction still earns more than momentum
($+9.2$ and $+33.0$ basis points unconditionally), but the high-$r$ filter
no longer separates good from bad periods. Three caveats limit how much
weight this result can bear: it is one panel out of eight, a $t$-statistic
of $2.1$ does not clear the higher hurdles appropriate after many tried
specifications \citep{harvey2016}, and the 24 coins share a common
market factor, which the portfolio construction absorbs but cannot
multiply into independent evidence. We therefore report it as a
conditional, venue-specific finding rather than a trading strategy.

\label{sec:results-bias}
The economic results also qualify the direction result for the naive
directional bias. For the deployed classifier, $\beta$ is weakly
\emph{negatively} rank-correlated with future direction in Binance
cryptocurrency (median IC $-0.03$ to $-0.05$, sign test $p<0.05$ at all
four horizons) and in FX at two horizons, and not significantly related
elsewhere; under opening-time alignment the same statistic had appeared
weakly positive, another instance of the timing bias of
Section~\ref{sec:results-forecast}. That the deployed classifier's phase
leans contrarian while the public classifier's phase earns a small
trend-following return in the same panel shows that whatever directional
information the phase carries depends on the classifier and is not a
robust property of coherence. The symmetry theorem of
Section~\ref{sec:methods-decomposition} explains why the odd component
of $\beta$ carries no information about even statistics of the path
under sign-symmetric returns. That theorem concerns independence from
volatility-type statistics, not unpredictability of direction, and the
cryptocurrency rule above shows that the sign of the coherence phase can
carry a small amount of directional information where returns are
trending. Everywhere else the direction result remains null.

\subsection{The variance ratio is doing real work too: it finds a random
walk, almost everywhere}
\label{sec:results-vr}

\begin{figure}[htbp]
\centering
\includegraphics[width=0.78\linewidth]{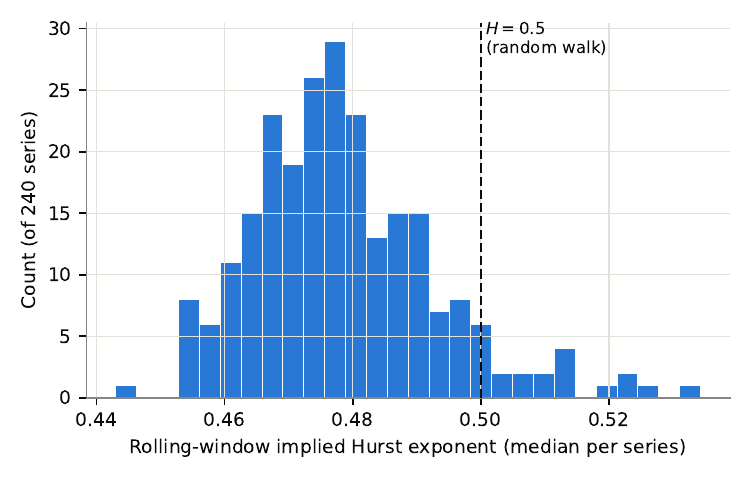}
\caption{Distribution of the rolling-window implied Hurst exponent, one
value per (source, instrument, timeframe) series ($n=240$), clustered
tightly around the random-walk value $H=0.5$.}
\label{fig:hurst-hist}
\end{figure}

Across all 240 (source, instrument, timeframe) combinations tested, the
whole-series variance-ratio verdict was \emph{random walk} for 191
(79.6\%), \emph{mean reverting} for 45 (18.75\%), and \emph{trending} for
only 4 (1.7\%). The rolling-window implementation (250-bar windows, 20-bar
steps) finds a median implied Hurst exponent of 0.477 (5th--95th
percentile 0.458--0.505, Figure~\ref{fig:hurst-hist}) and a median
rejection rate of 2.7\% -- below the estimator's own Monte-Carlo-calibrated
false-positive rate of 4.5--6.2\% (Sections~\ref{sec:methods-vr},~\ref{sec:validation}).

The rejection rate stays below the calibrated band by source (2.0\%
Binance, 3.3\% MT5, 2.35\% yfinance) and by timeframe (1.0\% at 1d to
3.85\% at 1m); 12.5\% of the 240 individual series do exceed 5\%, so this
is a strong central tendency, not an absolute statement about every
series.

\label{sec:results-discretization}
A brief caveat on the order parameter's own resolution: a previously
reported measurement found that a 4-timeframe ladder yields only 33
distinct values of $r$ across 13{,}000+ observations, and a 3-timeframe
ladder only 10 -- few timeframes means few achievable label combinations,
a property of the construction rather than of any one dataset. Because
this panel mixes 3-rung (yfinance) and 4-rung (all other sources) ladders,
a fixed numeric threshold would miscalibrate across it by construction;
percentile-based, per-instrument calibration avoids the problem but was
not independently re-verified for each of the eight new panels here
(future work).

\section{Discussion}
\label{sec:discussion}

Cross-timeframe phase coherence describes the recent past -- ``the market
just moved cleanly'' versus ``the market just chopped'' -- in a way that is
measurable and consistent across five very different asset classes. What
it does not do is forecast in any economically useful way outside one
market. Its rank correlation with the size of the next move is almost
entirely inherited from the persistence of volatility: once trailing
realized volatility and range are in the forecast, coherence adds a
median of zero to the out-of-sample $R^2$, and a fraction of a
percentage point only at short horizons in cryptocurrency. Its
directional content is likewise confined to cryptocurrency, where
trading the coherence phase at high coherence earns a gross return with
a break-even cost of about five basis points per round trip, enough for
maker execution on perpetual futures and too little for retail spot
trading. These are the results a finance reader should take from the
paper, and they sit comfortably with the adaptive-markets view
\citep{lo2004}: the most retail-driven, continuously traded and least
arbitraged market in the panel is the one where a multi-timeframe trend
signal survives, and it survives only at the margin of trading costs.
Separately, the variance ratio responsibly says only that, at the level
of raw returns, this panel is difficult to distinguish from a random
walk -- a narrow, well-defined claim about linear serial correlation,
not a claim that nothing is predictable by any method; the
cryptocurrency result is a concrete case of directional structure that
a non-rejection of the random walk does not rule out. Nor should the
two diagnostics be combined naively
as if correlated evidence for the same claim -- Section~\ref{sec:results-joint}
shows they are close to orthogonal, and
Sections~\ref{sec:methods-decomposition} and~\ref{sec:results-mechanism}
show this orthogonality is not an accident of this panel but a parity
property of the pair itself (an odd-dominated channel set against an
even statistic), so a researcher using both should
treat them as two independent pieces of evidence, not a two-vote
confirmation of one. The asset-class pattern is, however, an explanation
offered after the fact, not a hypothesis registered in advance, and
should be treated with corresponding caution.

\begin{figure}[htbp]
\centering
\includegraphics[width=0.78\linewidth]{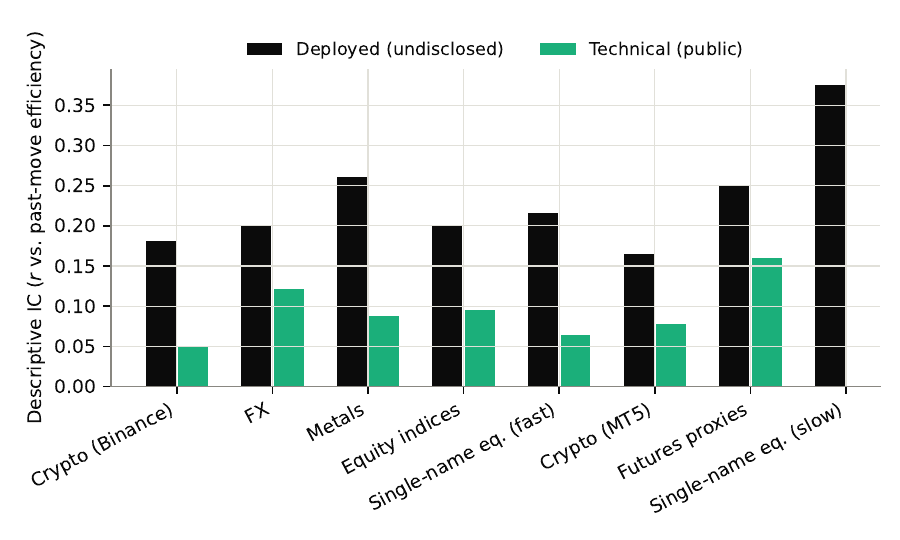}
\caption{Descriptive IC ($r$ vs.\ past-move efficiency) for the deployed
classifier against the public technical trend-volatility substitute
(rule-based, Equation~\ref{eq:technical-classifier}), by panel, both
with closed-bar information timing. The public classifier tracks the
deployed classifier's sign in every panel where it can be measured, at
reduced magnitude; the slow single-name equity panel has too few
non-overlapping origins for the public-classifier evaluation.}
\label{fig:classifier-comparison}
\end{figure}

These results come with real caveats, the most consequential of which is
that the regime classifier producing the discrete phase labels in
Section~\ref{sec:methods-coherence} is part of a deployed commercial trading-analytics
platform \citep{bunnyquant} and is not disclosed; it cannot be
independently re-implemented from this paper alone, even though the
Kuramoto mechanics downstream of classification are fully general and
reproducible with \emph{any} four-state regime classifier. To gauge how
much this matters in practice rather than leaving it as an abstract
caveat, we re-ran the full descriptive measurement of
Section~\ref{sec:results-descriptive} with the fully public technical
trend-volatility rule of Equation~\ref{eq:technical-classifier}
(Figure~\ref{fig:classifier-comparison}). It reproduces the direction and
sign of the descriptive finding in every panel with enough
non-overlapping observations to measure it, at roughly a quarter to
two-thirds of the deployed classifier's magnitude. We
read this as evidence that the descriptive finding is not an artifact
confined to one undisclosed classifier -- a simple, fully specified public
rule recovers the same qualitative pattern -- but also that the deployed
classifier captures real additional structure beyond what this simplest
public alternative sees. This paper does not attempt to resolve every
question that follows from that gap; a broader sweep of public
classifiers (statistical and learned, not just rule-based) and a
systematic sensitivity analysis is deliberately left to separate,
dedicated work, and only the one comparison load-bearing for this paper's
own reproducibility claim is reported here.

A second limitation concerns how far the permutation control reaches: the
shuffle-control evidence against the directional-bias artifact
(Sections~\ref{sec:validation},~\ref{sec:results-bias}) exists only for
the cryptocurrency panel, and the other seven panels' null results are
consistent with, but do not independently establish, the same mechanism --
re-running that control across the remaining panels is the obvious next
step before treating the direction-null result there as fully
established. Statistical power is also uneven across panels: metals
($n=2$) and futures proxies ($n=4$) are too small to support a confident
null, so their inclusion here is for completeness and transparency, not as
load-bearing evidence for any claim above. A related, structural
limitation is that three of the eight panels (yfinance-sourced, limited by
provider-side intraday history caps) use a 3-rung timeframe ladder against
the other five panels' 4-rung ladder; Section~\ref{sec:results-discretization}
explains why this affects the order parameter's discretization, but we
have not separately quantified how much of the cross-panel variation in
Tables~\ref{tab:descriptive} and~\ref{tab:forecast-oos} is attributable to ladder
length versus genuine asset-class differences, a natural target for
follow-up work once longer intraday history becomes available for those
instruments. Finally, with the exception of the two independently-sourced
cryptocurrency panels (Binance and the broker CFD feed, which agree in
sign and rough magnitude), each asset class here is observed through a
single provider, so venue-specific quoting or liquidity artifacts cannot
be ruled out as a partial driver of the cross-panel differences reported;
extending every panel to a second, independently-sourced venue, the way
the cryptocurrency panel already is, would be the most direct way to close
that gap. Finally, the symmetry theorem of
Section~\ref{sec:methods-decomposition} is exact but conditional on
sign-symmetric returns, and real markets violate that assumption to a
measurable degree -- returns are mildly skewed, and volatility responds
asymmetrically to down moves versus up moves (the leverage effect). This
is the most natural candidate explanation for why the real panel's
joint correlations (Table~\ref{tab:joint}) are small but not exactly
zero where the simulation's symmetric processes would predict zero;
quantifying how far empirically-calibrated asymmetry opens the direction
channel is a well-posed follow-up the decomposition now makes possible.

Measured identically across five asset classes and sixty-one instruments,
a Kuramoto-style cross-timeframe coherence measure reveals market
structure -- the directional cleanliness of recent price action -- that an
independently-calibrated variance-ratio test, applied to the same price
history, cannot detect: the two diagnostics are statistically independent
of each other rather than redundant restatements of one signal, a result
we did not anticipate going in and the one we consider the paper's main
contribution. That contribution rests on two legs rather than one:
the independence is measured across the panel
(Section~\ref{sec:results-joint}), and it is \emph{derived} -- an exact
decomposition splits the coherence measure into a direction channel that
a symmetry theorem keeps provably independent of the variance ratio, and
a volatility channel whose apparent coupling a controlled simulation
exposes as a dose-insensitive finite-sample artifact
(Sections~\ref{sec:methods-decomposition},~\ref{sec:results-mechanism}).
The finance benchmarks then set the limits of that structure: it adds
almost nothing to standard volatility forecasts, and its economic value
is confined to a gross edge in cryptocurrency that breaks even at about
five basis points per round trip
(Sections~\ref{sec:results-forecast},~\ref{sec:results-econ}).
We argue that stating plainly what a diagnostic cannot do,
and under what conditions its positive result does and does not hold, is
as important to report as the headline positive findings themselves.

The open question this leaves is not specific to markets. Wherever a
diagnostic imported from physics is set alongside a field's own classical
null-hypothesis test -- synchronization measures against
linear-correlation tests in neural or ecological time series, order
parameters against standard statistical-mechanics null models in any
other complex system with an established one -- the same question
applies, and the result reported here suggests the same answer may often
hold too: independence, not redundancy, between what the imported tool
sees and what the field's own test sees. The decomposition offered here
also suggests \emph{where to look first} in any such comparison: before
attributing a disagreement between two diagnostics to deep structure,
check their parity -- a tool with an odd-symmetric component set against
an even-symmetric test is guaranteed a degree of independence by symmetry
alone, in any system whose fluctuations are approximately
sign-symmetric, financial or otherwise. The most direct way to find out
whether that is a property of markets specifically or a property of this
kind of comparison in general is to run it once outside finance entirely,
on a complex system where a synchronization-style diagnostic and a
discipline-native null test both already exist separately but have, as
far as we are aware, never been compared on the same data -- the natural
next step this paper sets up but does not itself take.

\section*{Funding}
This research did not receive any specific grant from funding agencies in
the public, commercial, or not-for-profit sectors.

\section*{Declaration of competing interest}
The author is the founder of BunnyQuant Capital, a trading-analytics
platform. The regime classifier used for the panel-level independence and
descriptive measurements is proprietary to that platform and is not
disclosed. All forecast and economic results, the decomposition theorem
and the simulation study use the fully specified public classifier of
Equation~\ref{eq:technical-classifier} and can be reproduced from public
data and this paper.

\section*{Data availability}
Price data were obtained from public or retail sources (exchange APIs,
Yahoo Finance via the \texttt{yfinance} package, and a retail MetaTrader~5
feed); instrument lists and sample windows are given in
Section~\ref{sec:methods-data}. The public substitute classifier, the
coherence and variance-ratio code, and the simulation study are available
from the author upon reasonable request.

\section*{Declaration of generative AI and AI-assisted technologies in the manuscript preparation process}
During the preparation of this work the author used an AI language-model
assistant to help with drafting, editing and code support. After using this
tool, the author reviewed and edited the content as needed and takes full
responsibility for the content of the published article.

\bibliographystyle{elsarticle-num}

\end{document}